\documentclass[a4paper,fleqn,usenatbib]{mnras}

\usepackage{newtxtext,newtxmath}

\usepackage[T1]{fontenc}
\usepackage{ae,aecompl}

\usepackage{graphicx}	
\usepackage{amsmath}	
\usepackage{amssymb}	
\usepackage{multicol}        
\usepackage{bm}		
\usepackage{pdflscape}	

\title[Galactic Rotation with Cepheids]{Galactic Rotation from Cepheids with Gaia DR2 and Effects of Non-Axisymmetry}

\author[D. Kawata et al.]{
Daisuke Kawata$^{1}$\thanks{E-mail: d.kawata@ucl.ca.uk}, Jo Bovy$^{2}\thanks{Alfred P. Sloan Fellow}$, Noriyuki Matsunaga$^{3}$ and Junichi Baba$^{4}$
\\
$^{1}$Mullard Space Science Laboratory, University College London, Holmbury St. Mary, Dorking, Surrey, RH5 6NT, UK\\
$^{2}$Department of Astronomy and Astrophysics, University of Toronto, 50 St. George Street, Toronto, ON M5S 3H4, Canada\\
$^{3}$Department of Astronomy, The University of Tokyo, 7-3-1 Hongo, Bunkyo-ku, Tokyo 113-0033, Japan\\
$^{4}$National Astronomical Observatory of Japan, Mitaka, Tokyo 181-8588, Japan\\
}

\date{Accepted XXX. Received YYY; in original form ZZZ}

\pubyear{2017}

\begin{document}
\label{firstpage}
\pagerange{\pageref{firstpage}--\pageref{lastpage}}
\maketitle
%
\begin{abstract}
We apply a simple axisymmetric disc model to 218 Galactic Cepheids whose accurate measurements of the distance and velocities are obtained by cross-matching an existing Cepheids catalogue with the {\it Gaia} DR2 data.  Our model fit determines the ``local centrifugal speed'', $V_\mathrm{c}$---defined as the rotation speed required to balance the local radial gravitational force---at the Sun's location to be $V_{\rm c}(R_0)=236\pm 3$~km~s$^{-1}$ and the Sun's azimuthal and radial peculiar motions to be $V_{\rm \sun}=12.4\pm0.7$~km~s$^{-1}$ and  $U_{\rm \sun}=7.7\pm0.9$~km~s$^{-1}$, respectively. These results are obtained with strong priors on the solar radius, $R_0=8.2\pm0.1$~kpc, and Sun's angular rotation velocity, $\Omega_{\rm \sun}=30.24\pm0.12$~km~s$^{-1}$~kpc$^{-1}$. We also applied the axisymmetric model to mock data from an $N$-body/hydrodynamic simulation of a Milky Way-like galaxy with a bar and spiral arms. We find that our axisymmetric model fit to the young stars recovers the local centrifugal speed reasonably well, even in the face of significant non-axisymmetry. However, the local centrifugal speed determined from our Cepheid sample could suffer from systematic uncertainty as large as 6~km~s$^{-1}$.
\end{abstract}




\section{Introduction}
\label{sec:intro}

The circular velocity at the location of the Sun is a fundamental parameter in the determination of the distribution of the total mass in the Galaxy. However, measuring the circular velocity at the Sun is not a straightforward process, and different measurements provide systematically different results \citep[see][for a review]{bhg16}. The Sun's rotation velocity with respect to the Galactocentric rest frame, $\Omega_{\rm \sun}$, is most reliably measured as the proper motion of Sgr~A$^*$, assuming that Sgr~A$^{*}$ is fixed at the Galactic centre. Using Very Long Baseline Interferometry (VLBI), \citet{Reid+04} measured the proper motion of Sgr~A$^*$ with respect to a background quasar, and obtained $\Omega_{\rm \sun}=30.24\pm0.12$~km~s$^{-1}$~kpc$^{-1}$. However, the distance to the Galactic centre, $R_0$, and the difference between the circular velocity, $V_{\rm circ}(R_0)$, and the Sun's rotation speed, $V_{\phi, \sun}=\Omega_{\rm \sun} R_0$, remain challenging measurements. 
Various measurements of $R_0$ are converging to around $R_0=8.2\pm0.1$~kpc \citep{bhg16}. However, there are still systematic differences between the measurements \citep{rdggb15,bhg16}. The difference between $V_{\rm circ}(R_0)$ and $V_{\phi, \sun}$ is referred to as the Sun's peculiar rotation velocity, $V_{\sun}=V_{\phi, \sun}-V_{\rm circ}(R_0)$, and evaluated using different kinematic tracers combined with Galactic disc dynamical models. 

Star forming regions or young stellar populations are often used to perform this measurement, because their mean rotation velocity is believed to be close to the circular velocity, i.e. asymmetric drift can be ignored. For example, using the accurate astrometric measurements of about 100 high-mass star forming regions, \citet{rmbzd14} measured $V_{\rm \sun}=14.6\pm5.0$~km~s$^{-1}$ with $R_0=8.34\pm0.16$~kpc \citep[see also][]{Honma+2012}.

Classical Cepheids are also considered to be a good kinematic tracer for this purpose, because of their young age \citep[20$-$300~Myr, e.g.][]{Bono+05} and accurately-measured distance based on the well-known period-luminosity relation \citep{Inno+13}. \citet{Feast+97} combined the proper motions measured by {\it Hipparccos} and the photometric distances of 220 Galactic Cepheids, and determined $V_{\rm \sun}=11$~km~s$^{-1}$ and the angular velocity of circular rotation at the Sun, $\Omega_{\rm circ}=V_{\rm circ}(R_0)/R_0=27.19\pm0.87$~km~s$^{-1}$~kpc$^{-1}$ assuming $R_0=8.5\pm0.5$~kpc. Recently, \citet{Bobylev17a} combined the distance and line-of-sight velocity in \citet{mrbdr15} with the proper motions from the first {\it Gaia} data release \citep[{\it Gaia} DR1,][]{Gaia+Prusti16,Gaia+Brown16,llbhkhb16} for 249 Cepheids, and obtained $V_{\rm \sun}=11.73\pm0.77$~km~s$^{-1}$ and $\Omega_{\rm circ}=29.04\pm0.71$~km~s$^{-1}$ assuming $R_0=8.0$~kpc. 

 It is important to note that the kinematics of both star-forming regions and young stars like Cepheids is likely systematically affected by their environment, like the spiral arms which trigger star formation \citep[e.g.][]{bammsw09,Bovy09,McMillan10}. \citet{crs15} demonstrate that the rotation velocity deduced by the gas terminal velocities can be affected by the bar and spiral arms, which leads to incorrect determinations of the mean rotation velocity. The stars are also affected by the bar \citep[e.g.][]{wd00}, spiral arms \citep[e.g.][]{jsjb02,DeSimone+Wu+Tremaine04,gkc12a,gkc12b,bsw13}, and possibly by the combination of bar and spiral arms \citep[e.g.][]{Quillen03,avpmff09,imbf10}. \citet{tbdr17} study how the Galactic parameters they determine with an advanced, action-based Galactic disc dynamical model are affected by the spiral arms \citep[see also][]{chszck16}. They conclude that the dynamical models can be biased due to the influence of spiral arms, but if the sample of stars covers a large enough volume (with a radius of at least 3 kpc), the dynamical model can recover the global Galactic parameters.  
 
 The {\it Gaia} mission \citep{Gaia+Prusti16} has made its second data release \citep[{\it Gaia} DR2][]{Gaia+Brown+18}, which provides accurately measured astrometry \citep{Lindegren+18} and line-of-sight velocities \citep{Cropper+18,Katz+RV+18,Sartoretti+18} for stars covering a sufficiently large volume around the Sun. However, as described in detail in \citet{Lindegren+18,Arenou+18}, the zero-point calibration of the parallax measurements is still preliminary in {\it Gaia} DR2, and systematic uncertainties are reported by comparing to other distance measurements, including those from the Cepheid period-luminosity relations \citep{Riess+18}. Therefore, to use {\it Gaia} DR2 parallax measurements for dynamical models, one must carefully consider the uncertainties. Also, the line-of-sight velocities in {\it Gaia} DR2 are median values of the measurements at the various epochs, which do not provide an accurate line-of-sight velocity for variable stars in their rest frame. Hence, in this paper, we employ the well calibrated Cepheids distances and line-of-sight velocities from the literature and combine them with the proper motion measurements from {\it Gaia} DR2. We apply the simple axisymmetric dynamical model used in \citet{baabbdc12} to this sample of classical Cepheids and deduce the Galactic parameters.
  
 
 We modify the axisymmetric kinematic model of \citet{baabbdc12} to fit both the line-of-sight velocity and the tangential velocity of stars, taking into account the observational uncertainties in distance and velocity.  This model takes into account the velocity dispersion and asymmetric drift, which  \citet{Bobylev17a} do not.  Hence, this model can also be applied for other kinds of stellar tracer populations. We further discuss how the deduced Galactic parameters are affected by non-axisymmetric structures and possible systematic motions of young stars, using an $N$-body/Smoothed Particle Hydrodynamics (SPH) simulation of a Milky Way-like disc galaxy with a bar and spiral arms. We demonstrate that young stars within 3~kpc from the observer form a suitable sample to recover the ``local centrifugal speed'', $V_{\rm c}$, which is defined in this paper as the rotation speed required to balance  the local radial gravitational force, which can be different from the azimuthally averaged global circular velocity at the radius.
  
 Section~\ref{sec:data} discusses the sample of Cepheids used for our axisymmetric disc model fit, which itself is described in Section~\ref{sec:axisymdm}. The results of the axisymmetric disc model fit to the Cepheids data are shown in Section~\ref{sec:Ceph-res}. Section~\ref{sec:mocktest} demonstrates that the axisymmetric model can recover the model parameters well for mock data created with the axisymmetric disc with known parameters, as expected. Section~\ref{sec:simdata} discusses the results of applying the axisymmetric model to the $N$-body/SPH simulation data of a Milky Way-like disc galaxy. A summary and discussion of this study are presented in Section~\ref{sec:sum}.

\section{Cepheids data}
\label{sec:data}

 Our sample is taken from \citet{glbrf14}, where distances to Cepheids are determined homogeneously by using near-infrared data. Uncertainties in distance modulus are $0.05-0.07$~mag for most of the Cepheids. We then cross-match this sample with the {\it Gaia} DR2  proper motion data \citep{Lindegren+18}. We further cross-match with a sample of Cepheids whose line-of-sight velocity, $V_{\rm los}^*$, are provided in \citet{mrbdr15}. We use $V_{\rm los}^*$ here to indicate the full line-of-sight velocity to distinguish it from its projection onto the disc plane, which is what we use below. For cross-matching these catalogues, we used \textsc{topcat} \citep{mbt05}. The cross-match provides 274 Cepheids with known locations and 3D kinematics.  We further limit the sample based on vertical position with respect to the Sun, $|\tilde{z}|<0.2$ kpc. To take into account the error, we define $\tilde{z}=\sin({b})10^{(DM+DM_{\rm err}-10)/5.0}$~kpc, where $b$, $DM$, and $DM_{\rm err}$ are Galactic latitude, distance modulus and uncertainty in the distance modulus, respectively. This limit was applied to eliminate clear outliers, because our sample shows a clear concentration around the Galactic plane, with about 70~\% of the Cepheids located within 100~pc of the mid-plane, as expected for young stars like Cepheids. To avoid having the results of our modelling be affected by a few Cepheids at large distances with larger relative velocity with respect to the Sun, we also limit the sample to a distance of 4~kpc, beyond which the number of Cepheids drops dramatically. These selection cuts leave 218 Cepheids in our sample.

\begin{figure}
\includegraphics[width=\hsize]{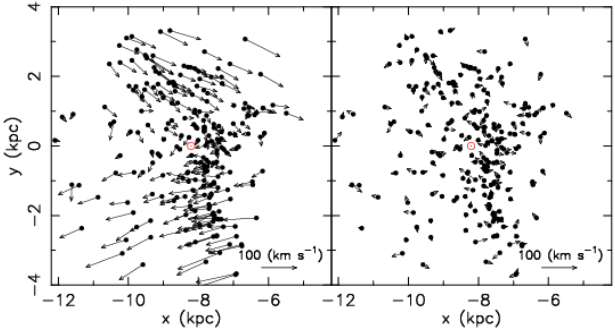}
\caption{Distribution of the Cepheids used for our model fit. Arrows in the left (right) panel indicate their heliocentric velocity  before (after) subtracting the mean rotation of the best-fit axisymmetric model.}
\label{fig:galmap}
\end{figure}

 In this paper, we ignore the vertical motion or the thickness of the disc, and only consider the distance and velocity projected onto the disc plane. Hence, our axisymmetric model introduced in the next section is a two-dimensional model in the disc plane. Therefore, we use the heliocentric velocities projected onto the disc plane, i.e. $V_{\rm los}=V_{\rm los}^* \cos(b)$, where $V_{\rm los}^*$ is the full line-of-sight velocity. We evaluate the uncertainty in the two-dimensional velocity, $\sigma_{V_{\rm glon}}$ and $\sigma_{V_{\rm los}}$, where $\sigma_{V_{\rm glon}}$ and $\sigma_{V_{\rm los}}$ are the uncertainties of the velocity measurements in the direction of Galactic longitude, $V_{\rm glon}$, and heliocentric line-of-sight velocity projected on the disc plane, $V_{\rm los}$, respectively. $\sigma_{V_{\rm glon}}$ was evaluated by taking the standard deviation of a Monte-Carlo (MC) sampling of $V_{\rm glon}$ computed from randomly selected right ascension (RA) and declination (Dec) proper motions, and distance taken from randomly selected distance modulus. The proper motions were sampled using their 2D Gaussian probability distributions with their measured means, uncertainties and the correlation between the RA and Dec proper motions. The distance modulus was taken from the Gaussian probability distribution with the measured distance modulus as the mean and their 1~$\sigma$ uncertainty as the standard deviation. For most of the Cepheids included, all of these two velocity errors are less than 4~km\,s$^{-1}$ and are roughly comparable to each other. The distribution of the Cepheids sample is shown in Fig.~\ref{fig:galmap}. The arrows in the left panel of the figure show the velocities of Cepheid with respect to the Sun. 

\section{Model}
\label{sec:axisymdm}

\subsection{Axisymmetric disc kinematic model}

Following \citet{baabbdc12}, we compute the mean and dispersion of $V_{\rm los}$ expected in an axisymmetric Galactic disc model in the Galactic rest frame, and derive the posterior probability for the parameters of the models by exploring likelihood of the model parameters with Markov-Chain Monte-Carlo (MCMC). The model is axisymmetric and assumes a Gaussian velocity distribution in the directions radial and tangential to Galactic rotation with no correlation between these components, and zero mean radial velocity.  Obviously, these are considered to be simple assumptions to describe the kinematics of the Galactic disc. However, in Section~\ref{sec:simdata} we demonstrate that these simple approximations are sufficient to determine the local kinematic parameters by applying the model to young stars in a relatively small region of the disc. Compared to \citet{baabbdc12} who only used $V_{\rm los}$, we include the component of the velocity in the direction of Galactic longitude, $V_{\rm glon}$. Because both $V_{\rm los}$ and $V_{\rm glon}$ contribute to rotational and radial velocities, we take into account the covariance in the likelihood function, as shown in Section~\ref{sec:mcmc_param}. 

In the axisymmetric disc model, we only need to consider the mean rotation velocity, $\overline{V_{\phi}}$, and velocity dispersion in the Galactic rotation, $\sigma_{\phi}$, and radial, $\sigma_{R}$, directions, because the mean radial velocity and the correlation between the two velocity components are zero. The mean rotation velocity is calculated from asymmetric drift, $V_{\rm a}$, as $\overline{V_{\phi}}(R)=V_{\rm circ} (R)-V_{\rm a}(R)$, where $V_{\rm circ}(R)$ is the circular velocity at the radius $R$. Following \citet{baabbdc12}, we assume an exponential disc (stellar density $\propto \exp[-R/h_{R}]$) and an exponentially-declining radial velocity dispersion ($\sigma_{R}\propto \exp[-R/h_{\sigma}]$). Then the asymmetric drift can be calculated from the Jeans equation
\begin{equation}
V_{\rm a}(R)=\frac{\sigma_{R}^2(R)}{2 V_{\rm c}(R)} \left[(\sigma_{\phi}/\sigma_{R})^2-1+R\left(\frac{1}{h_R}+\frac{2}{h_{\sigma}}\right)\right],
\label{eq:vasym}
\end{equation}
\citep[e.g.][]{Binney+Tremaine08}. We assume that $(\sigma_{\phi}/\sigma_{R})^2$ is constant for simplicity. As discussed in \citet{baabbdc12}, we confirmed that the Galactic parameters we are interested in are not sensitive to the radial scale length,  $h_R$, of radial density profile or the radial scale length , $h_{\sigma}$, of radial velocity dispersion. Here, we fix $h_R=20$~kpc and $h_{\sigma}=20$~kpc. These values are inspired from the scale lengths seen for the star particles in a similar age range as the Cepheids in our numerical simulation described in Section~\ref{sec:simdata}. The assumed $h_R$ is much larger than the conventional value for the Milky Way thin disc, e.g. typical values of $h_R=2.5$ to $5$~kpc found by \citet{brlhbl12}. However, the scale length for the young stars around the Sun is not well known, and could be different from that for the old stars. For example, \citet{Mackereth+Bovy+Schiavon17} analysed the stellar structure of the Galactic disc as a function of age using data from the Apache Point Observatory Galactic Evolution Experiment (APOGEE) data, and found that the younger stars have a flatter or even peaked radial profiles around the solar radius. The scale length of radial velocity dispersion profile is also not well constrained. From the kinematics of old disc K giants, \citet{Lewis+Freeman89} showed $h_{\sigma}=4.37$~kpc. \citet{Huang+Liu+Yuan16} used red clump stars from APOGEE and the LAMOST Spectroscopic Survey of the Galactic Anti-centre and suggested $h_{\sigma}=16.40\pm1.25$~kpc. The scale length of the radial velocity dispersion profile for young stars could be different from those of relatively older stars. Hence, we use the values obtained from our numerical simulation mentioned above.  As a result, our axisymmetric model can be described with $V_{\rm circ}(R)=\overline{V_{\phi}}(R)+V_{\rm a}(R)$, $\sigma_{R}(R)$ and $(\sigma_{\phi}/\sigma_{R})^2$. We also assume that $V_{\rm circ}$ follows a linear function of $R$ with the slope of $d V_{\rm circ}(R_0)/dR$ within the radial range of our sample. As mentioned above, the velocity dispersion profile is assumed to be an exponential. We parameterize the amplitude of the radial velocity dispersion profile with $\sigma_{R}(R_0)$, the radial velocity dispersion at $R_0$. Finally, $R_0$ is also a free parameter in our analysis. 

In the Galactic rest frame, the mean rotational velocity at the position of the star, $\overline{V_{\phi}}$, can be projected onto the line-of-sight velocity, $V_{\rm los}$, from the observer, i.e. the position of the Sun, as $V_{\rm m,los}=\overline{V_{\phi}} \sin(\phi+l)$, where $l$ is Galactic longitude and $\phi$ is the angle between the line from the Galactic centre toward Sun and the one toward the position of the star, positive in the direction of Galactic rotation. 
Following the same strategy, we can derive the projected longitudinal velocity from the rotation velocity of the axisymmetric model as  $V_{\rm m, glon}=\overline{V_{\phi}}\cos(\phi+l)$.

The observational data provide the line-of-sight velocity, $V_{\rm los}^{\rm helio}$, and Galactic longitudinal velocity, $V_{\rm glon}^{\rm helio}$, with respect to the Sun. Using the solar radial velocity, $V_{R,\odot}$ (outward motion is positive) and rotational velocity, $V_{\rm \phi,\odot}$ (clock-wise rotation is positive, $V_{\rm \phi,\odot}=V_{\rm circ}(R_0)+V_{\rm \odot}$) in the Galactic rest frame, these velocities can be converted to the Galactocentric rest frame as follows.
\begin{eqnarray}
V_{\rm o,los} & = & V_{\rm los}^{\rm helio} - V_{R,\odot} \cos l + V_{\rm \phi,\odot} \sin l, \\
V_{\rm o,glon} & = & V_{\rm glon}^{\rm helio} + V_{R,\odot} \sin l + V_{\rm \phi,\odot} \cos l.
\end{eqnarray}
These values can be compared with the expected velocity distribution of $V_{\rm m, los}$ and $V_{\rm m, glon}$ from the model. Hence, fitting our axisymmetric disc model to the observed data, $V_{\rm los}^{\rm helio}$ and $V_{\rm glon}^{\rm helio}$ for the tracer sample, requires us to explore 7 model parameters, $\theta_{\rm m}=\{V_{\rm circ}(R_0), V_\mathrm{\sun}, V_{R,\odot}, \sigma_{R}(R_0), (\sigma_{\phi}/\sigma_{R})^2, R_0, dV_{\rm circ}(R_0)/dR\}$. We choose $V_\mathrm{\sun}=V_{\phi, \sun}-V_\mathrm{circ}(R_0)$ as a parameter rather than $V_{\phi, \sun}$, because we find that our Cepheids sample provides stronger constraints on the peculiar rotational velocity of the Sun. This is due to the small velocity dispersion and negligible asymmetric drift of the sample of Cepheids. In fact, our result in Section~\ref{sec:Ceph-res} indicates a very small asymmetric drift of $V_\mathrm{a}(R_0)=0.28\pm0.2$~km~s$^{-1}$ at $R_0$ for Cepheids.

\subsection{MCMC parameter probabilities}
\label{sec:mcmc_param}

We use Bayes' theorem to find the marginalised probability distribution function of our model parameters as follows.
\begin{equation}
 p(\theta_{\rm m}|\mathcal{D}) =  \mathcal L(\mathcal{D}|\theta_{\rm m}) \times Prior\,,
\end{equation}
where $\mathcal{D}=\left(V_{\rm los}^{\rm helio}, V_{\rm glon}^{\rm helio}\right)$ denotes the whole set of the observed values for all stars in our sample, and $\theta_{\rm m}$ denotes the combination of all of the model parameters.
 We run MCMC using this $p(\theta_{\rm m}|\mathcal{D})$.
The likelihood function is
\begin{equation}
\mathcal{L}(\mathcal{D}| \theta_{\rm m})=\prod_i^N \frac{1}{2 \pi |\mathbfss{C}_i|^{1/2}} 
 \exp\left(-0.5  \mathbfit{X}^{\rm T}_i  \mathbfss{C}_i^{-1}  \mathbfit{X}_i \right), 
\end{equation}
where
\begin{equation}
\mathbfit{X}_i= \left(
 \begin{array}{c}
 V_{\rm o,los,i}-V_{\rm m,los,i} \\
 V_{\rm o,glon,i}-V_{\rm m,glon,i} 
 \end{array} \right).
\end{equation}
Here, $V_{\rm o,los,i}$ and $V_{\rm o,glon,i}$ are the observed line-of-sight and longitudinal velocity in the Galactic rest frame for star $i$, and $V_{\rm m,los,i}$ and $V_{\rm m,glon,i}$ are the expected line-of-sight and longitudinal velocity in the Galactic rest frame from the axisymmetric model at the location of star $i$. The covariance matrix, $ \mathbfss{C}_i$, is $ \mathbfss{C}_i=  \mathbfss{A}_i  \mathbfss{S}_i  \mathbfss{A}_i^{\rm T}$, where
\begin{equation}
 \mathbfss{A}_i= \left(
 \begin{array}{cc}
  -\cos(\phi+l) & \sin(\phi+l) \\
  \sin(\phi+l) & \cos(\phi+l) 
 \end{array} \right),
\end{equation}
and
\begin{equation}
 \mathbfss{S}_i= \left(
 \begin{array}{cc}
  \sigma_{R}^2 & 0 \\
  0 & \sigma_{\phi}^2
 \end{array} \right).
\end{equation}

To take the observational uncertainties in the distance modulus, line-of-sight velocity and proper motions into account, we average each Cepheid's likelihood over 1,000 MC samples from the uncertainty distribution of the observed distance modulus, line-of-sight velocity and RA and Dec proper motions from Gaussian distributions with their observed values and errors. When sampling the uncertainty distribution of the proper motions in RA and Dec, we take the correlation of the RA and Dec proper motion in the {\it Gaia} data into account.
This can be described as
\begin{equation}
 \mathcal{L}(\mathcal{D}| \theta_{\rm m})=\prod_i^N \int p(\mathcal{D}_{\rm obs,i} | \mathcal{D}_{\rm true,i}) \mathcal{L}(\mathcal{D}_{\rm true,i} | \theta_{\rm m}) d\mathcal{D}_{\rm true,i},
 \label{eq-mclike}
\end{equation}
where $\mathcal{D}_{\rm obs,i}$ are the observed values for the $i$-th Cepheid, and $\mathcal{D}_{\rm true,i}$ are the error-free, true values of these observables predicted by the model. We approximate the integral of equation (\ref{eq-mclike}) with the MC sampling as described above.

We find that $R_0$ is not well constrained by our data (see Section \ref{sec:Ceph-res}). Hence, we introduce a Gaussian prior for $R_0$ as follows.
\begin{equation}
 Prior(R_0)= \frac{1}{\sqrt{2 \pi \sigma^2_{R_{\rm 0,prior}}}}  \exp\left(-\frac{(R_0-R_{\rm 0,prior})^2}{2 \sigma_{R_{\rm 0,prior}}^2}\right),
\end{equation}
where we set $R_{\rm 0,prior}=8.2$ kpc and $\sigma_{R_{\rm 0,prior}}=0.1$ kpc from \citet{bhg16}. We further use a Gaussian prior for the angular velocity of the Sun, $\Omega_{\sun}=V_{\rm \phi,\odot}/R_0$, with $\Omega_{\sun}=30.24\pm0.12$ km s$^{-1}$ kpc$^{-1}$, because this is well constrained by \citet{Reid+04}.

We use the \texttt{emcee}  \citep{Goodman+Weare10,fmhlg13} MCMC sampler with 128 walkers and 1,000 chains per walker. We use \texttt{galpy} \citep{jb15} for coordinate transformations.

\begin{figure*}
\includegraphics[width=\hsize]{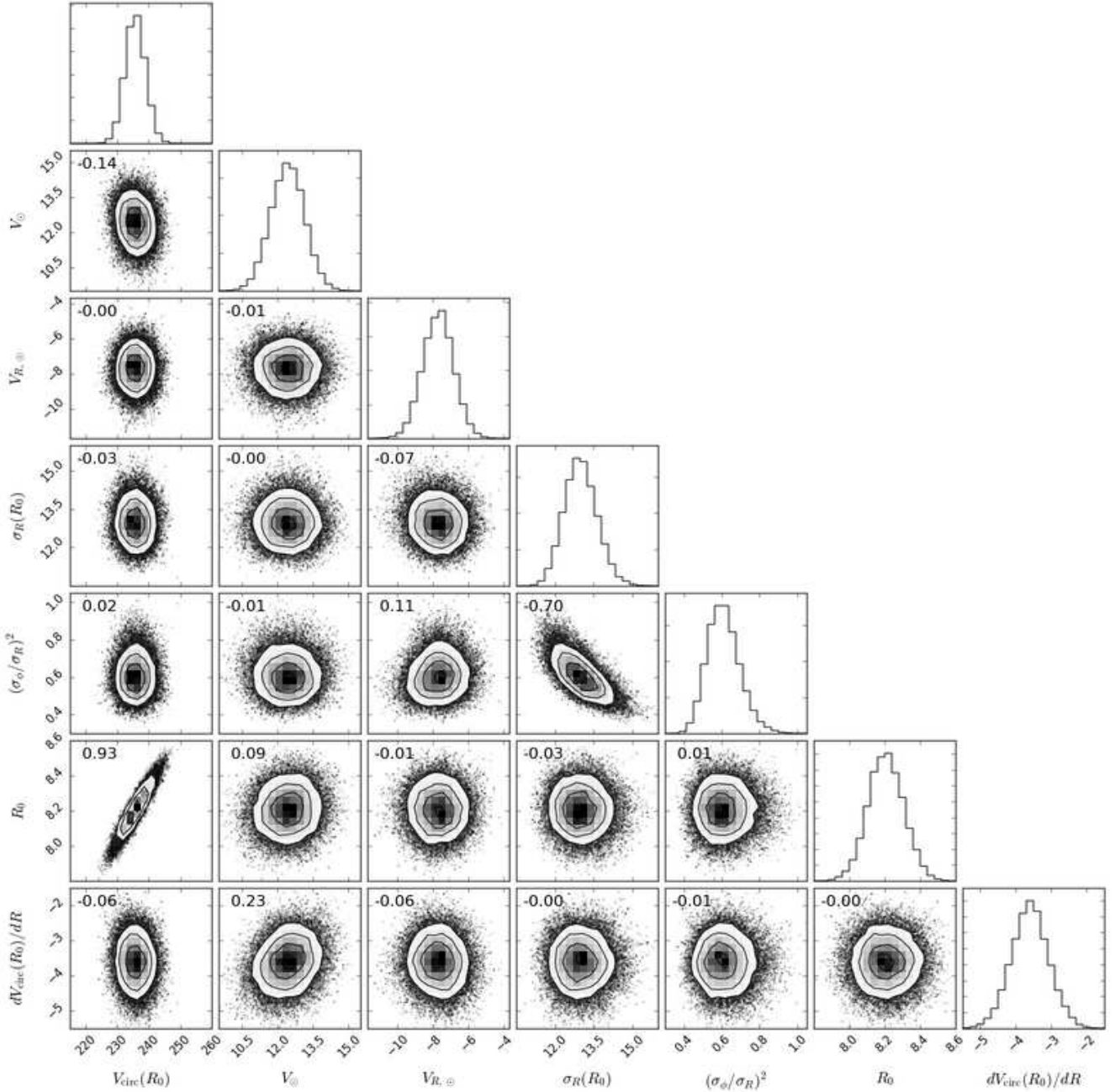}
 \caption{Marginalised posterior probability distribution of our model parameters in our fiducial model. The number in each panel gives the correlation coefficient.
 }
    \label{fig:ROMGprior-MCMC}
\end{figure*}

\begin{figure}
\includegraphics[width=\hsize]{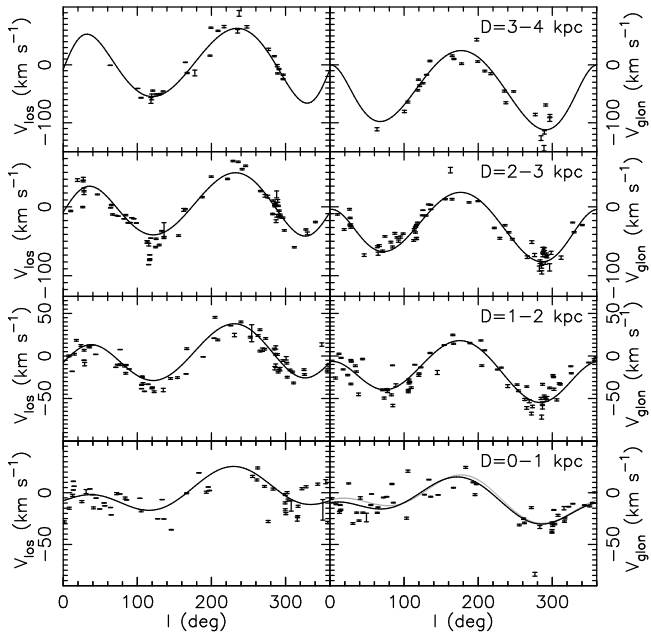}
 \caption{Line-of-sight velocity, $V_{\rm los}^{\rm helio}$, (left) and Galactic longitudinal velocity, $V_{\rm glon}^{\rm helio}$, (right) as a function of Galactic longitude for our sample of Cepheids (error bars) with distances between 3 and 4~kpc (top), 2 and 3~kpc (2nd), 1 and 2~kpc (3rd) and 0 and 1~ kpc (bottom). Many samples have uncertainties smaller than the thickness of the error bars. The solid line in each panel shows the result from our best fit model using the mean distance of the data in each panel. The grey line in the bottom right panel displays $V_{\rm glon}^{\rm helio}$ from the Oort constants $A$, $B$ and $C$ determined in \citet{jb17a}. The uncertainty is as small as the thickness of the grey line.}
    \label{fig:glonv}
\end{figure}

\begin{figure*}
\includegraphics[width=\hsize]{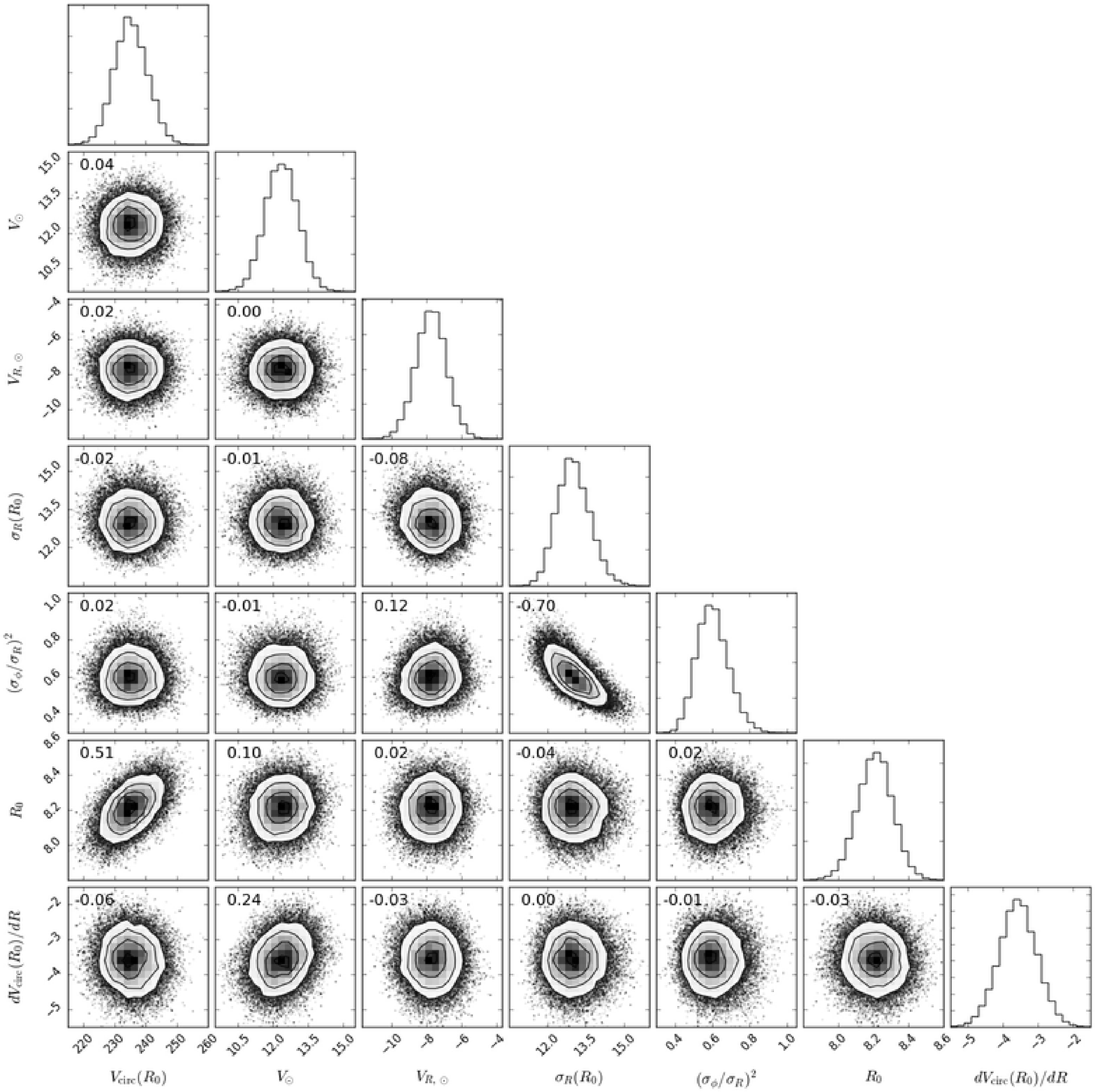}
 \caption{Marginalised posterior probability distribution of our model parameters in the model with only the $R_0=8.2\pm0.1$~kpc prior. 
The number in each panel gives the correlation coefficient. 
Note that the $x$- and $y$-axes ranges in each panel are different from Fig.~\ref{fig:ROMGprior-MCMC}.}
\label{fig:Rprior-MCMC}
\end{figure*}

\begin{table*}
 \caption{Results of the MCMC fit}
  \label{tab:MCMC-res}
 \begin{tabular}{lccccc}
  \hline
                                        & Fiducial$^{a}$                & $R_0=8.2\pm0.1^{b}$  &  $R_0=8.2\pm0.4$ 
  & $h_R=2.6$~kpc         &   \citet{Bobylev17a}$^{e}$ \\
  & & & $\Omega_{\sun}=30.24\pm0.12^{c}$
  & $h_{\sigma}=4.37$~kpc$^{d}$ & \\
 \hline
 $V_{\rm c}(R_0)$ (km~s$^{-1}$)  & $236\pm 3$     & $235\pm 5$                 & $240\pm10$ 
 &   $339\pm3$              & $231\pm6$ \\
 
 $V_{\sun}$ (km~s$^{-1}$)  & $12.4\pm0.7$            &  $12.4\pm0.7$             & $12.4\pm0.7$ 
 & $10.0\pm0.9$            &  $11.7\pm0.8$ \\

 $V_{R,\sun}$   (km~s$^{-1}$)   & $ -7.7\pm0.9$      & $-7.8\pm  0.9$             & $-7.7\pm0.9$ 
 & $-7.9\pm0.9$             &  $-7.9\pm0.7$ \\

 $\sigma_{R}(R_0)$  (km~s$^{-1}$) & $ 13.0\pm 0.6$  & $ 13.0\pm   0.7$  &  $13.0\pm0.6$  
 & $14.9\pm0.7$            &  $-$ \\

 $(\sigma_{\phi}/\sigma_{R})^2$  &  $  0.61\pm  0.09$  & $  0.60\pm   0.08$  & $0.61\pm0.08$ 
  & $0.67\pm0.10$         &  $-$ \\

 $R_0$  (kpc)                 &  $  8.2\pm   0.1$             & $  8.2\pm   0.1$           & $8.34\pm0.4$  
  & $8.2\pm0.1$             &  $8.0\pm0.2^{f}$  \\
  
 $dV_{\rm circ}/dR$  (km~s$^{-1}$~kpc$^{-1}$)   
                                      &  $ -3.6\pm   0.5$             & $ -3.6\pm   0.5$          &  $-3.6\pm0.5$ 
 & $-3.7\pm0.5$            & $-3.6\pm1.7^{g}$ \\

 $N$                                      & 218 &  218 & 218 & 218 & 249 \\
 
\hline
\end{tabular}

\begin{flushleft}
$^{a}$ Gaussian priors of $R_0=8.2\pm0.1$~kpc and $\Omega_{\sun}=30.24\pm0.12$ km s$^{-1}$ kpc$^{-1}$ are applied.\\
$^{b}$ Only a Gaussian prior of $R_0=8.2\pm0.1$~kpc is applied, and no $\Omega_{\sun}$ prior is used.\\
$^{c}$ Gaussian priors of $R_0=8.2\pm0.4$~kpc and $\Omega_{\sun}=30.24\pm0.12$ km s$^{-1}$ kpc$^{-1}$ are applied.\\
$^{d}$ Same as the fiducial case, but using $h_R=2.6$~kpc and $h_{\sigma}=4.37$~kpc.\\
$^{e}$ Results from \citet{Bobylev17a}.\\
$^{f}$ This is fixed.\\
$^{g}$ This is calculated from $R_0=8.0\pm0.2$~kpc, $\Omega_{\rm circ}=28.84\pm0.33$~km~s$^{-1}$~kpc$^{-1}$ and $d\Omega_{\rm circ}/dR=-4.05\pm0.10$~km~s$^{-1}$~kpc$^{-2}$.
\end{flushleft}
\end{table*}

\section{Results for Cepheids in {\it Gaia} DR2}
\label{sec:Ceph-res}

We apply our axisymmetric model fit described in Section~\ref{sec:axisymdm} to the sample of Cepheids shown in Fig.~\ref{fig:galmap}. The marginalised posterior probability distributions of the model parameters are shown in Fig.~\ref{fig:ROMGprior-MCMC}, and the mean and dispersion from each parameter's probability distribution are summarised in Table~\ref{tab:MCMC-res}. We refer to the result from this model as the  `fiducial model result'. Below, we relax the priors to study the sensitivity of our results to our prior assumptions.

The fiducial model results are similar to those of \citet{Bobylev17a} who applied a different method to a similar set of Cepheids data, but using {\it Gaia} DR1 \citep{Gaia+Brown16}. It is interesting that the radial velocity of the Sun deduced from  the kinematics of the Cepheids tends to provide a lower value ($V_{R,\sun}=-7.7\pm0.9$~km~s$^{-1}$ in our fiducial model result) compared to the ones from kinematics of red giants (e.g. $V_{R,\sun}=-10\pm1$~km~s$^{-1}$; \citealt{baabbdc12}) or local dwarf stars (e.g. $V_{R,\sun}=-11.1^{+0.69}_{-0.75}$~km~s$^{-1}$; \citealt{sbd10}). In addition to the parameters explored in \citet{Bobylev17a}, our model fit provides the velocity dispersion, $\sigma_{R}$, and the ratio of the azimuthal and radial velocity dispersion, $(\sigma_{\phi}/\sigma_{R})^2$, for the Cepheids sample. 

Arrows in the right panel of Fig.~\ref{fig:galmap} show the velocity of the Cepheids in the Galactocentric rest frame after subtracting the rotational velocity from the best-fit axisymmetric model. There are no obvious systematics in the residual velocity field which implies that the axisymmetric model is a reasonable model to describe the average velocity trend in our sample. As studied in \citet{bkmgh18}, the residual velocity fields are likely due to the influence of spiral arms and therefore provide valuable information about the nature of the spiral arms. Because of the relatively large area covered by our sample of Cepheids, our model fit is not so sensitive to the systematic velocity trend around the spiral arms \citep[see also][]{tbdr17}. Fig.~\ref{fig:glonv} compares the $V_{\rm los}^{\rm helio}$ and $V_{\rm glon}^{\rm helio}$ trends with Galactic longitude from the best fit model at different distance with the observed Cepheids kinematics. The figure demonstrates that the best fit model describes both velocity components of the Cepheids reasonably well at the different distances. The grey line in the bottom right panel in Fig.~\ref{fig:glonv} shows $V_{\rm los}^{\rm helio}$ obtained from the Oort constants, $A$, $B$ and $C$, determined in \citet{jb17a} from the {\it Gaia} DR1 proper motion data of nearby dwarf stars with a typical distance of 230~pc. The grey line is broadly consistent with our data within the distance of 1~kpc, and with our best fit model.


We obtain $V_\mathrm{\sun}=12.4\pm0.7$~km~s$^{-1}$ in our fiducial model result as shown in Table~\ref{tab:MCMC-res}. This is consistent with $V_\mathrm{\sun}=12.24\pm0.47$~km~s$^{-1}$ in \citet{sbd10} from the {\it Hipparcos} data of nearby (distance$<\sim200$~pc) dwarf stars. However, this is significantly different from $V_\mathrm{\sun}=26\pm3$~km~s$^{-1}$ in \citet{baabbdc12} who applied the same axisymmetric model to the APOGEE survey data which covers distances up to 10~kpc \citep[see also][who obtained similar values from the mean velocity field of the APOGEE red clump stars]{bbgmnz15}. As discussed in \citet{baabbdc12}, this may suggest that nearby stars covered by the {\it Hipparcos} stars have a systematic motion as large as 10~km~s$^{-1}$ with respect to the local standard of rest, due to a streaming motion caused by non-axisymmetric structures. Our sample covers distances up to $\sim4$~kpc, although the majority of stars are within 3~kpc. It is interesting to find that our results are still consistent with $V_\mathrm{\sun}$ deduced with the solar neighbourhood sample within the very small volume covered by the {\it Hipparcos} stars. However, the size of our sample is small, and we do not think that our result is conclusive enough to test the scenario of the streaming motion of the solar neighbourhood sample. Streaming motions and incomplete phase mixing are now observed in {\it Gaia} DR2 \citep{Antoja+18,Gaia+Katz18Disc,Kawata+18b}. How these motions affect the deduced $V_\mathrm{\sun}$ from the solar neighbourhood star sample should be tested with a more advanced dynamical model and careful consideration of systematic uncertainties in {\it Gaia} DR2.

Fig.~\ref{fig:ROMGprior-MCMC} presents the correlation coefficient of every combination of two parameters at the top-left corner in each panel. $V_\mathrm{circ}(R_0)$ is strongly correlated with $R_0$. To determine the effect of our prior, we ran the axisymmetric disc model fit without the prior on $\Omega_{\sun}$. The result is shown in Fig.~\ref{fig:Rprior-MCMC} and Table~\ref{tab:MCMC-res}. Comparison between  Figs.~\ref{fig:ROMGprior-MCMC} and \ref{fig:Rprior-MCMC} shows a weaker correlation of $V_\mathrm{circ}(R_0)$ with $R_0$ in the $R_0$-prior-only model (Fig.~\ref{fig:Rprior-MCMC}). Hence, the strong correlation between $V_\mathrm{circ}(R_0)$ with $R_0$ is due to our strong prior on the angular velocity of the Sun, $\Omega_{\sun}$.

Table~\ref{tab:MCMC-res} also shows the results of a model with a prior on $R_0$ of $R_0=8.2\pm0.4$ kpc and the same $\Omega_{\sun}$ prior as the fiducial model. This prior prefers a higher value of $R_0$. This indicates that our Cepheids data poorly constrain the Galactocentric radius of the Sun. 

As discussed above, the scale lengths $h_R$ and $h_{\sigma}$ for the young stellar population, like Cepheids, are not well known. We assumed a large $h_R=20$~kpc and $h_{\sigma}=20$~kpc, motivated from our numerical simulation. To estimate the systematic uncertainty due to this assumption, we ran a model with much smaller $h_R=2.6$~kpc \citep{bhg16} and $h_{\sigma}=4.37$~kpc \citep{Lewis+Freeman89}, as measured for the older stellar population. Table~\ref{tab:MCMC-res} shows the result of this model. Because of the larger asymmetric drift expected from equation~(\ref{eq:vasym}), the deduced $V_\mathrm{circ}(R_0)$ becomes higher, and therefore $V_{\sun}$ becomes smaller. This provides a systematic uncertainty of about 3~km~s$^{-1}$ due to the uncertainty in $h_R$ and $h_{\sigma}$.

\begin{figure*}
\includegraphics[width=\hsize]{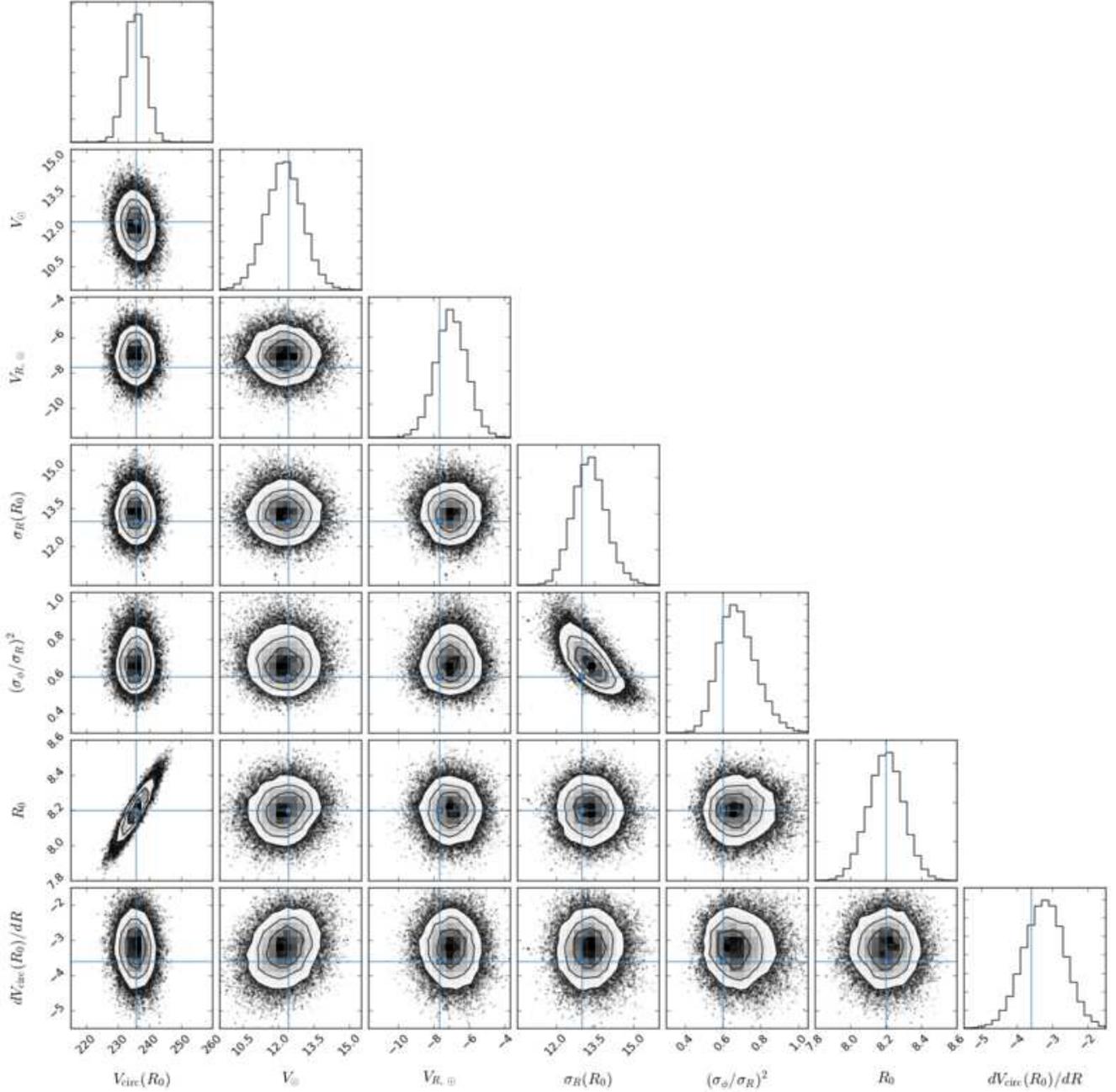}
 \caption{Marginalised posterior probability distribution of the MCMC fit to the axisymmetric disc mock data. The lines indicate the input parameters.}
    \label{fig:axdmock-MCMC}
\end{figure*}

\begin{table*}
 \caption{Results of the MCMC fit to axisymmetric disc mock data}
 \label{tab:axdmockMCMC-res}
 \begin{tabular}{lcccc}
  \hline
                                        & Input     & Model & $DM_\mathrm{sys}=-0.1$~mag & $DM_\mathrm{sys}=+0.1$~mag\\
 \hline
 $V_{\rm circ}(R_0)$ (km~s$^{-1}$)                  & 236         & $236\pm 3$     &  $236\pm3$       &  $235\pm3$ \\
 $V_{\sun}$ (km~s$^{-1}$)                                & 12.4        & $12.3\pm 0.8$  &  $12.2\pm0.7$  &  $12.3\pm0.8$ \\
 $V_{R,\sun}$ (km~s$^{-1}$)                            & $-7.7$     & $ -7.0\pm 0.9$  &  $-6.7\pm0.9$   &  $-7.4\pm0.9$ \\
 $\sigma_{R}(R_0)$ (km~s$^{-1}$)                   & 13.0 & $ 13.3\pm 0.6$ &  $12.9\pm0.6$  &  $13.7\pm0.6$ \\
 $(\sigma_{\phi}/\sigma_{R})^2$                        &  0.6 & $  0.7\pm 0.1$   &  $0.7\pm 0.1$   & $0.7\pm0.1$ \\
 $R_0$ (kpc)                                                      & 8.2          & $  8.2\pm 0.1$  &  $8.2\pm0.1$    & $8.2\pm0.1$ \\
 $dV_{\rm circ}/dR$ (km~s$^{-1}$~kpc$^{-1}$) &  $-3.6$    & $ -3.2\pm 0.6$  &  $-3.6\pm0.6$   & $-2.9\pm0.6$ \\
\hline
\end{tabular}
\end{table*}

\section{Axisymmetric disc mock data test}
\label{sec:mocktest}

To validate our MCMC fitting algorithm, we make mock data using the position of the 218 observed Cepheids and assign a velocity expected at the location of the Cepheids from an axisymmetric disc model with known input parameters. Then, we create a mock data set by randomly displacing the distance modulus, the line-of-sight radial velocity and the proper motions using the observed uncertainties of each Cepheid at each location. We perform the axisymmetric model fit to the mock data, taking into account the observed uncertainty with MC sampling as described in Section~\ref{sec:axisymdm}. We use the same $R_0$ and $\Omega_{\sun}$ priors as our fiducial model fit in Section~\ref{sec:Ceph-res}.

Marginalised posterior probability distribution of the parameters are shown in Fig.~\ref{fig:axdmock-MCMC} and the results are summarised in Table~\ref{tab:axdmockMCMC-res}. Table~\ref{tab:axdmockMCMC-res} also shows the input parameters used for creating the mock data. All parameters are recovered well within their 1~$\sigma$ uncertainty. The uncertainties from the model fit to the mock data are also similar to those in the fiducial model results for the real  data in Table~\ref{tab:MCMC-res}. This validates that our method provides the expected quality of the results from the number and accuracy of the distances and velocities in our Cepheids sample. The strong correlation between $V_\mathrm{circ}(R_0)$ and $R_0$ is also seen in the mock data results in Fig.~\ref{fig:axdmock-MCMC}.

The distance modulus measured for Cepheids could be affected by systematic uncertainties of about 0.1~mag \citep{Inno+13}. To evaluate the effect of such systematic uncertainties on our results, we added a systematic offset of $DM_\mathrm{sys}=+0.1$~mag and of $DM_\mathrm{sys}=-0.1$~mag to the distance modulus of the mock data and then fit without taking this offset into account. Our axisymmetric model fit results in these two cases are shown in Table~\ref{tab:axdmockMCMC-res}, where the $DM_\mathrm{sys}=-0.1$ ($+0.1$)~mag case corresponds to the result for the mock data whose distance moduli were decreased (increased) by 0.1~mag. Table~\ref{tab:axdmockMCMC-res} shows that all parameters are still recovered reasonably well. 
There are some systematic trends, because the offset $DM_\mathrm{sys}=-0.1$ ($+0.1$) leads to underestimated (overestimated) distance and proper motion. $V_{R,\sun}$ and $\sigma_{R}(R_0)$ become smaller (larger) for $DM_\mathrm{sys}=-0.1$ ($+0.1$), which means that the radial velocity is more sensitive to the proper motion, compared to the line-of-sight velocity. Also, $dV_{\rm circ}/dR$ becomes higher (lower) for $DM_\mathrm{sys}=-0.1$ ($+0.1$), because the distance is underestimated (overestimated). Still, the results provide a reasonable recovery, and this demonstrates that a systematic distance uncertainty of 0.1~mag does not affect our results significantly. 

\begin{figure}
\includegraphics[width=\hsize]{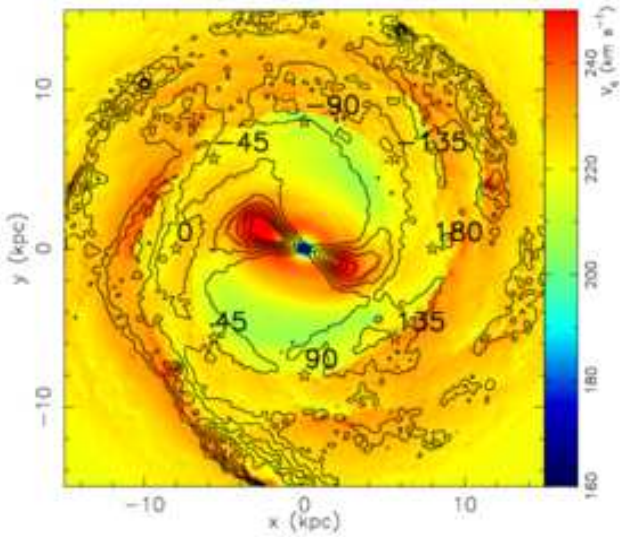}
 \caption{Map of the local centrifugal velocity, $V_\mathrm{c}(x,y)$, defined in equation (\ref{eq:Vcircxy}), at different positions in the disc over-plotted with contours of the over-density of stars. The positions indicated with the empty star symbols are the location of the observers chosen for our mock data. The numbers indicate the angles in Fig.~\ref{fig:vcvsigcomp}.}
    \label{fig:vcmap}
\end{figure}

\section{Fitting mock data from an $N$-body/SPH Milky Way-like simulation}
\label{sec:simdata}

The Galactic disc is not purely axisymmetric. The motions of stars are affected by non-axisymmetric structures, such as the bar and spiral arms \citep[e.g.][]{lbk72,wd00,bammsw09,gkc12a,gkc14,dmhcss13,bbgmnz15}. Gravitational forces in the disc plane are also affected by the non-axisymmetric structures. Therefore, $V_\mathrm{circ}$ obtained using our axisymmetric model fit to the local Cepheids sample is likely not exactly the circular velocity, but the rotation velocity required to explain the local kinematics, when it is assumed to be a part of an axisymmetric disc. In other words, this rotation velocity is a rotation speed required to balance the local radial gravitational force. We call this rotation velocity the ``local centrifugal speed'', which can be described as
\begin{equation}
V_\mathrm{c}(x,y)=[F_\mathrm{R}(x,y) R(x,y)]^{1/2},
\label{eq:Vcircxy}
\end{equation}
where $F_\mathrm{R}(x,y)$ is the gravitational force in the radial direction toward the Galactic centre at the position of $(x,y)$ in the Galactic plane. The local centrifugal speed can be different at the different azimuthal angles even at the same Galactocentric radius \citep{crs15,chszck16,tbdr17}. 

In this section, we discuss what the local centrifugal speed obtained by our axisymmetric modelling for the Cepheids kinematics means, applying the axisymmetric model fit to particle data from a numerically simulated disc galaxy similar to the Milky Way. We use the same snapshot data as the DYN model at $t=2.62$~Gyr in \citet{bkmgh18}. This is a Milky Way-like disc simulation with an $N$-body/SPH code, {\tt ASURA}, with self-gravity, radiative cooling, star formation, and stellar feedback \citep{sdk08,sm09,sm10}. This particular snapshot shows a clear bar component and $m=2$ spiral arms, and \citet{bkmgh18} showed that the motion of young star particles around a Perseus-like spiral arm in the simulation is similar to what is observed for the \citet{glbrf14} Cepheids sample with the Gaia DR1 proper motions. 

\begin{figure}
\centering
\includegraphics[width=\hsize]{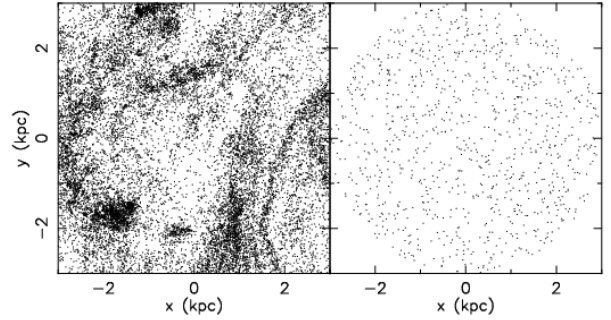}
 \caption{Distribution of all (left panel) and selected (right panel) star particles whose age is between 20 and 300 Myr. The x-y coordinate is the same as Fig~\ref{fig:vcmap}, and the panels are centred at the assumed solar location, $(x,y)=(-8,0)$.}
    \label{fig:partsel}
\end{figure}

\begin{figure}
\centering
\includegraphics[width=\hsize]{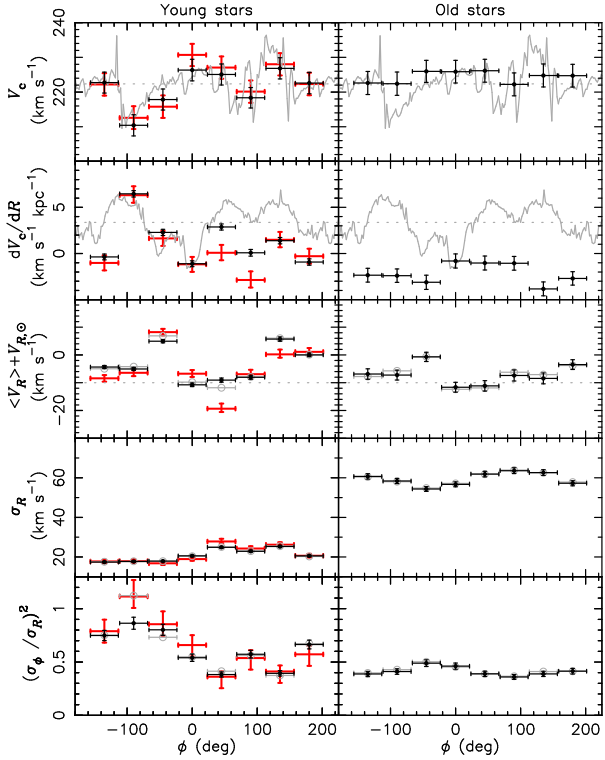}
 \caption{Black filled circles with vertical error bars show the results from our model fit for the young (left) and old (right) star particles as a function of the azimuthal angle of the assumed observer position.  Red thick error bars indicate the fit results for the mock Cepheid data (see the main text for more detail). Grey solid lines indicate the true values from the simulation model, and horizontal dashed lines indicate the mean of these values in the top and 2nd panels.  In the 3rd panel, the grey circles indicate the assumed observer's radial motion (grey dotted line) plus the mean radial velocity of the sampled particles. In the 4th and bottom panels, grey open circles with the horizontal error bars show the true values.}
    \label{fig:vcvsigcomp}
\end{figure}

Fig.~\ref{fig:vcmap} shows the local centrifugal speed at different positions around the disc and the contours show the over-density of the stars, which clearly show the position of the bar and spiral arms.  Fig.~\ref{fig:vcmap} shows that the local centrifugal speed changes significantly around the bar and spiral arms. Hence, the local centrifugal speed at different azimuthal positions can be different even at a fixed radius. The observer position used in \citet{bkmgh18} corresponds to the position of $(x,y)=(-8,0)$ kpc in Fig.~\ref{fig:vcmap}, and we define this to correspond to $\phi=0$~deg. Open star symbols in Fig.~\ref{fig:vcmap} show the position at different $\phi$ at $R=8$~kpc, and $\phi$ increases in the anti-clockwise direction. The grey line in the top panels in Fig.~\ref{fig:vcvsigcomp} indicates the local centrifugal speed as a function of $\phi$ at $R=8$~kpc. The horizontal grey dotted line shows the azimuthally averaged centrifugal speed at $R=8$~kpc. 
We can see that the local centrifugal speed can be different by about 10~km~s$^{-1}$ from the averaged centrifugal speed.

We created mock Cepheids data from the simulation snapshot by selecting the star particles whose age is between 0.02 and 0.3~Gyr similar to the estimated age range of Cepheids \citep[e.g.][]{Bono+05}. At $\phi=0$~deg, these young star particles are distributed as shown in the left panel of Fig.~\ref{fig:partsel}. It is obvious that the spatial distribution of the young stars is not homogeneous. Instead, the particles are concentrated around the spiral arms, and there are several clumps with different sizes. We applied our axisymmetric model fit to these particle data, but because of this biased sampling around the spiral arm, the model fit could not recover the true local centrifugal speed. To mitigate this problem, we have selected 1,000 particles randomly, but as homogeneous as possible, which are shown in the right panel of Fig.~\ref{fig:partsel}. This is further motivated by the fact that stars in our Cepheids data sample are distributed reasonably homogeneously, as seen in Fig.~\ref{fig:galmap}. To select these particles, we randomly selected a location within the distance of 3~kpc from the observer position, and picked the closest particle from the randomly selected location after eliminating the particles already picked up. We have chosen a distance limit of 3~kpc, because the majority of the Cepheids data in Fig.~\ref{fig:galmap} are within 3~kpc from the Sun, and we found that the larger distance limit breaks the axisymmetric assumption and provides worse results. We limited the number of particles to 1,000, because the particle resolution of the simulation allows us to pick the young star particles randomly and reasonably homogeneously up to around this number. We also assumed that the observer at $\phi=0$~deg and $R_0=8$~kpc was moving with $V_{\phi,\sun}=235$~km~s$^{-1}$ and $V_{R,\sun}=-10$~km~s$^{-1}$. In our axisymmetric model fit, we assume $h_R=20$~kpc and $h_{\sigma}=20$~kpc in equation~(\ref{eq:vasym}), because this roughly matched the density and velocity dispersion profile. As mentioned above, we used the same $h_R$ and $h_{\sigma}$ for our model fit in Section~\ref{sec:Ceph-res}. We also apply the priors of $R_{0,\rm prior}=8.0\pm0.1$~kpc and $\Omega_{\sun, \rm prior}=235.0/8.0\pm0.12$ ~km~s$^{-1}$~kpc$^{-1}$ similar to what we did in Section~\ref{sec:axisymdm}.

The results of the axisymmetric model fit for these selected young particles around the assumed observer position of $(x,y)=(-8,0)$ are shown at $\phi=0$ deg in the left panel of Fig.~\ref{fig:vcvsigcomp} (black filled circle with error bars). The top left panel shows that the true local $V_\mathrm{c}$ at $\phi=0$ (grey solid line) is slightly higher than the mean centrifugal speed (grey dotted line), and our axisymmetric model fit (black filled circle) recovers this local centrifugal speed well. More remarkably, the left panel in the second row shows that the gradient of $V_{\rm c}$ at $\phi=0$ is recovered very well, although the true local gradient shown with the solid grey line is significantly different from the azimuthally averaged $dV_\mathrm{c}/dR$ (horizontal grey dotted line). 
The left panel in the third row shows the result for the radial velocity of the observer, $V_{R,\sun}$, for which the true value is $-10$~km~s$^{-1}$ (grey dotted line). Our model fit assumed that the mean radial velocity of the sample, $\langle V_{R}\rangle$, is zero, but if there is non-zero $\langle V_{R}\rangle$, the fit result provides $\langle V_{R}\rangle+V_{R,\sun}$. Therefore, this panel shows the true $\langle V_{R}\rangle+V_{R,\sun}=\langle V_{R}\rangle-10$~km~s$^{-1}$ from the simulation data as an open grey circle with horizontal error bar. Our model fit result is shown with black filled circle with error bars. The mean radial velocity of the $\phi=0$~deg sample happened to be close to zero, and our model fit result recovered the assumed $V_{R,\sun}$ very well. The left panel in the fourth row shows the result of $\sigma_{R}$ for the sampled star particles. The measured $\sigma_{R}$ from the simulation data is shown with an open grey circle with horizontal error bars. Our model fit result is shown with a black filled circle with error bars (vertical error bars are not visible, because the inferred uncertainty in $\sigma_{R}$ from our model fit is very small), which reproduces the true velocity dispersion very well. The square of velocity dispersion ratio, $(\sigma_{\phi}/\sigma_{R})^2$, is also recovered well at $\phi=0$~deg in the bottom left panel. Again, the true $(\sigma_{\phi}/\sigma_{R})^2$ measured directly from the particle data is shown with grey open circles with horizontal error bars, and the model fit results are shown with black filled circles with error bars. 

Because the local centrifugal speed depends on the azimuthal angle even at a fixed radius as shown in the top left panel of Fig.~\ref{fig:vcvsigcomp}, we apply the same exercise at different azimuthal positions at $R=8$~kpc, as indicated with open star symbols in Fig.~\ref{fig:vcmap}. We sampled 1,000 young star particles with age between 20 and 300 Myr randomly and spatially homogeneously within a distance of 3 kpc from the observer after placing the observer position at different $\phi$. Then, we apply our axisymmetric model fit to the sample, assuming the same observer's rotation speed of $V_{\phi,\sun}=235$~km~s$^{-1}$ and radial motion of $V_{R,\sun}=-10$~km~s$^{-1}$. The results and comparison with the true values from the simulation data are shown in the left panels of Fig.~\ref{fig:vcvsigcomp}. Interestingly, the local centrifugal speed, $V_\mathrm{c}$, is well recovered at the different $\phi$. The deduced local centrifugal speed is underestimated at $\phi=-45$ and $-90$~deg, but considering the angle covered by the tracer sample, it is not much more than a $1\sigma$ difference. The successful recovery of the local centrifugal speed at different $\phi$ in this complex simulation with our simple axisymmetric model is encouraging. This means that applying an axisymmetric model to a relatively small region of the disc area can be valid to recover the local centrifugal speed.

The left panel in the second row shows that the radial gradient of centrifugal speed, $dV_\mathrm{c}/dR$, is also recovered reasonably well at $\phi\leq0$~deg, while it is struggling at $\phi\geq90$~deg. We find that the recovery of $dV_{\rm circ}/dR$ is worse, where $V_{\rm circ}$ changes non-monotonically, and shows a significant oscillation within $\pm3$~kpc range in radius, because of the spiral arms. The rest of panels show the reasonable agreement between true values and deduced values. However, $(\sigma_{\phi}/\sigma_{R})^2$ is significantly lower at $\phi=-90$~deg, where the spiral arm is affecting the velocity distribution. Still, the local centrifugal speed and the gradient is recovered well at $ \phi=-90$~deg. This also indicates that our axisymmetric model fit provides a more robust estimate of the local centrifugal speed than of the other parameter values. 

Interestingly, for the Sun-like location, $\phi=0$, all values are very well recovered. This location is between the two strong spiral arms, like the Perseus and Scutum-Centaurus arms in the Milky Way. The simple axisymmetric model works reasonably well at such a location. 

We made more comparable mock data to our Cepheids data by sampling the position and velocities of the young ($20<$ Age $<300$~Myr) star particles nearest to the observed position of 218 Cepheids used in Section~\ref{sec:Ceph-res} after placing the observer position at different $\phi$ and $R_0=8$~kpc. We then applied our axisymmetric disc model to the mock data, taking into account the same observational errors as the Cepheids data closest to the corresponding particle. The results are shown with red open square with the error bars in the left panels of Fig.~\ref{fig:vcvsigcomp}. The results show larger errors than when we use all the particles with no errors (black filled circle).  The local centrifugal speed is overestimated at $\phi=0$~deg by 6.2~km~s$^{-1}$, which is larger than the statistical uncertainties of 3.2~km~s$^{-1}$. This indicates that caution is necessary when interpreting the results from our Cepheid sample results. Still, our model fit results recover the true parameter values within $1\sigma$ uncertainty or slightly more. In fact, Fig.~\ref{fig:galmap} shows that our Cepheids sample are distributed reasonably randomly, which must also help our good recovery of the true parameter values. 

We further perform the same exercise for a sample of 1,000 stars selected randomly in the same way as we did for the young star particles, but for star particles whose age is between 3.5 and 5 Gyr. Note that the simulation snapshot used is at $t=2.62$~Gyr. In the initial condition, the formation time, $t_{\rm form}$, of the star particles are randomly assigned between $-5$ and 0 Gyr, and the age was calculated by age $=2.62-t_{\rm form}$~Gyr. Hence, these particles are star particles that are part of the initial condition with the axisymmetric particle density distribution. Also, note that as shown later, the radial velocity dispersion of this old star sample is about 60~km~s$^{-1}$, which is as high as the thick disc population in the Milky Way. Therefore, the selected star particles do not constitute a representative sample of stars with a similar age in the Milky Way, but they rather represent the older stars similar in age to the oldest stars in the Milky Way disc. Hence, we simply call this sample ``old stars''.

The results for the old stars are shown in the right panels of Fig.~\ref{fig:vcvsigcomp}. As in the left panels, the true values from the simulations are shown with grey lines or symbols, and our model fit results are shown as black filled circles with error bars. The model fit results show less variations with $\phi$. For the local centrifugal speed, the model fit results for the old stars provide a similar local centrifugal speed to the average centrifugal speed or slightly higher than that. The results are not sensitive to the fluctuations of the local centrifugal speed with $\phi$, because of the large velocity dispersion of the old stars. 
These results indicate that the old stars with large velocity dispersion are a good tracer to recover the azimuthally averaged properties of the disc, and that they are not too sensitive to the local perturbations like spiral arms. However, the gradient of the centrifugal speed is completely off from the mean and from the local gradient values, except at $\phi=0$~deg. This may not be surprising, because as shown in the top right panel of Fig.~\ref{fig:vcvsigcomp}, the old stars' kinematics are insensitive to the variation in the local centrifugal speed.

Even for the old stars, the mean radial velocity can be non-zero at some locations in the disc, which are indicated by the difference between the grey open circle and the horizontal dashed line in the plot of $\langle V_{R}\rangle+V_{R,\sun}$. This means that unless $\langle V_{R}\rangle$ is independently measured by another method, the local velocity distribution fit only provides $\langle V_{R}\rangle+V_{R,\sun}$ and it is difficult to recover the true $V_{R,\sun}$ in the Galactocentric rest frame. 

From the results of this section, we conclude that young stars are good tracers for the axisymmetric model fit to recover the local centrifugal speed at different locations in the Galactic disc. Cepheid variables provide the ideal sample of young stars within a tightly constrained age range. However, their number density is not large and it would be difficult to increase the sample size by a factor of 5 or more, as used in this section. Hence, we think that the young dwarf stars would be a good tracer population for this purpose. The {\it Gaia} DR2 provides accurate distances and proper motions for many young dwarf stars. For example, A-type dwarf stars are expected to be younger than 1~Gyr, and have a small enough velocity dispersion for this application. Applying the axisymmetric disc model locally to the A-type dwarf stars in {\it Gaia} DR2 would be an alternative way of recovering the local centrifugal speed at the solar position and also at different Galactocentric radii and azimuths. Unfortunately, the line-of-sight velocity, $V_{\rm los}^{\rm helio}$, of the A-type stars are not available in the {\it Gaia} DR2 data, because of the high effective temperature of A-type stars. Therefore, this would require us to combine the {\it Gaia} DR2 parallax and proper motion measurements with existing spectroscopic survey data or apply the axisymmetric model to only the longitudinal velocities, $V_{\rm glon}^{\rm helio}$. 

\section{Summary and Discussion}
\label{sec:sum}

We determined the local centrifugal speed, $V_\mathrm{c}$, defined as a rotation speed required to balance the local radial gravitational force, and its radial gradient from 218 Galactic Cepheids whose accurate measurements of the distance and velocities are obtained by cross-matching the existing Cepheids catalogue with the {\it Gaia} DR2 data. Our axisymmetric disc model fit also provides the Sun's radial and azimuthal velocity, and the velocity dispersion of the sample of stars. 

Assuming strong priors on the solar radius, $R_0=8.2\pm0.1$~kpc, and on the angular rotation velocity of the Sun, $\Omega_{\rm \sun}=30.24\pm0.12$~km~s$^{-1}$~kpc$^{-1}$, we have obtained $V_\mathrm{c}(R_0)=236\pm 3$~km~s$^{-1}$, $V_\mathrm{\sun}=V_{\rm \phi,\sun}-V_{\rm circ}(R_0)=12.4\pm0.7$~km~s$^{-1}$, $V_{R,\sun}=-7.7\pm0.9$~km~s$^{-1}$, $\sigma_{R}=13.0\pm0.6$~km~s$^{-1}$, $(\sigma_{\rm \phi}/\sigma_{R})^2=0.61\pm 0.09$ and $dV_{\rm circ}/dR=-3.5\pm0.5$~km~s$^{-1}$~kpc$^{-1}$. Here, $V_{R,\sun}$ is positive in the outward direction. For the more conventional definition of the solar radial motion, $U_{\sun}$, whose positive direction is inward, this is equivalent to $U_{\sun}=7.7\pm0.9$~km~s$^{-1}$. These results are consistent with those of \citet{Bobylev17a} who used a similar set of Cepheids data, but a different model fit. Our model provides $\sigma_{R}$ and $(\sigma_{\rm \phi}/\sigma_{R})^2$, which are not fitted in \citet{Bobylev17a}.

Because this circular velocity is deduced with modelling of a relatively local sample of young disc stars, we call this the `local centrifugal speed' in this paper. This speed is defined as $V_\mathrm{c}(x,y)=[F_\mathrm{R}(x,y) R(x,y)]^{1/2}$, where $F_\mathrm{R}(x,y)$ is the radial gravitational force at the position $(x,y)$ in the Galactic plane. Using an $N$-body/SPH simulation of a Milky Way-like disc galaxy and locating a mock observer at different azimuthal locations around the Galactic disc at a fixed radius similar to the solar radius, we demonstrate that our axisymmetric model fit to the sample of young stars within 3~kpc from the observer recovers the local centrifugal speed at the location of the observer well. This can be different from the azimuthally averaged centrifugal speed at the radius, because the bar and spiral arms lead to different radial forces, $F_\mathrm{R}(x,y)$, at different azimuthal locations even at the same Galactocentric radius. The local centrifugal speed and the other Galactic parameters are recovered well, only when we used young star particles whose age is as young as classical Cepheids, and also when we selected the particles randomly in a spatially-homogeneously manner. We also find that the local centrifugal speed deduced from our Cepheid sample could suffer from a systematic uncertainty as large as 6~km~s$^{-1}$, which is larger than the statistical uncertainty. 

This is an encouraging result and suggests that the simple axisymmetric model approximation is valid when using the sample of young stars within a small volume with a radius of 3~kpc to deduce the local centrifugal speed. The model fit result at different locations in the Galactic disc can thus provide the variation of the local centrifugal speed, which would be valuable information to understand the influence of the bar and spiral arm on the local gravitational field and stellar motions. 


The final data release of the {\it Gaia} data will provide accurate positions and velocities for different kinds of stellar populations in a large volume of the Galactic disc. It is promising that advanced Galactic dynamical modelling \citep[e.g.][]{Bovy+Rix13,sb15,Trick+Bovy+Rix16} will uncover the circular velocity profile of the azimuthally averaged Galaxy. $N$-body based Galaxy modelling can take into account non-axisymmetric structures and is expected to reveal the structure of the bar and possibly the spiral arms \citep{Hunt+Kawata+Martel13,Hunt+Kawata14,Portail17}. We think that applying the simple axisymmetric model to different regions of the Galactic disc in the {\it Gaia} data would be a simpler alternative way of providing the local centrifugal speed and its radial gradient. The model is simpler than the advanced methods listed above. However, a simpler model is often useful to validate a more complicated advanced model. Hence, we believe that this model would be also useful for studying the properties of the Galactic disc with the {\it Gaia} data. This paper also demonstrates that applying the model to mock data from Milky Way-like simulated galaxies is an important ingredient to understanding the limits of dynamical modelling. 
 



 \section*{Acknowledgments}
We thank an anonymous referee for their thorough review and helpful suggestions which have improved the manuscript significantly. Kawata acknowledges the generous support and hospitality of the National Astronomical Observatory of Japan. 
Kawata also acknowledges the support of the UK's Science \& Technology Facilities Council (STFC Grant ST/N000811/1). Bovy received partial support from the Natural Sciences and Engineering Research Council of Canada. Bovy also received partial support from an Alfred P. Sloan Fellowship. Matsunaga and Baba acknowledge the Japan Society for the Promotion of Science (JSPS) Grant-in-Aid, KAKENHI, No. 18H01248. Baba is supported by the JSPS Grant-in-Aid for Young Scientists (B) Grant Number 26800099 and for Scientific Research (C) Grant Number 18K03711. The $N$-body/SPH simulation used in this paper were carried out on facilities of Center for Computational Astrophysics (CfCA), National Astronomical Observatory of Japan. This work also used the UCL facility Grace and the DiRAC Data Analytic system at the University of Cambridge, operated by the University of Cambridge High Performance Computing Service on behalf of the STFC DiRAC HPC Facility (www.dirac.ac.uk). This equipment was funded by BIS National E-infrastructure capital grant (ST/K001590/1), STFC capital grants ST/H008861/1 and ST/H00887X/1, and STFC DiRAC Operations grant ST/K00333X/1. DiRAC is part of the National E-Infrastructure. This work has made use of data from the European Space Agency (ESA) mission Gaia (https://www.cosmos.esa.int/gaia), processed by the Gaia Data Processing and Analysis Consortium (DPAC, https://www.cosmos.esa.int/web/gaia/dpac/consortium). Funding for the DPAC has been provided by national institutions, in particular the institutions participating in the Gaia Multilateral Agreement.   



\bibliographystyle{mnras}
\bibliography{./dkref} 







\bsp	
\label{lastpage}
\end{document}